# Characterization of pixel crosstalk and impact of Bayer patterning by quantum efficiency measurement

*Jérôme Vaillant[a], Clémence Mornet[b,c], Thomas Decroux[b], Didier Hérault[b], Isabelle Schanen[c]*

[a]STMicroelectronics, 12 rue Jules Horowitz, BP 217, 38019 Grenoble Cedex, FRANCE;
[b]STMicroelectronics, 850 rue Jean Monnet, 38926 Crolles Cedex, FRANCE;
[c]IMEP, 3 rue Parvis Louis Néel, BP 257, 38016 Grenoble Cedex 1, FRANCE

## ABSTRACT

Development of small pixels for high resolution image sensors implies a lot of challenges. A high level of performance should be guaranteed whereas the overall size must be reduced and so the degree of freedom in design and process. One key parameter of this constant improvement is the knowledge and the control of the crosstalk between pixels. In this paper, we present an advance in crosstalk characterization method based on the design of specific color patterns and the measurement of quantum efficiency. In a first part, we describe the color patterns designed to isolate one pixel or to simulate un-patterned colored pixels. These patterns have been implemented on test-chip and characterized. The second part deals with the characterization setup for quantum efficiency. Indeed, the use of spectral measurements allows us to discriminate pixels based on the color filter placed on top of them and to probe the crosstalk as a function of the depth in silicon, thanks to the photon absorption length variation with the wavelength. In the last part, results are presented showing the impact of color filters patterning, *i.e.* pixels in a Bayer pattern versus un-patterned pixels. The crosstalk directions and amplitudes are also analyzed in relation to pixel layout.

## 1. INTRODUCTION

Small pixel is a key enabler for a lot of image sensor market.[1–4] The main one is obviously the mobile phone market and its demand for compact high resolution sensors. Nevertheless, small pixels are interesting for several fields like webcam, digital still camera, medical application, etc. Nevertheless, the development of these pixels faces a lot of challenges due to the very limited space available for transistors and interconnections but also due to proximity effect between pixels. One key parameter of pixel quality is the control of the crosstalk,[5–7] i.e. the unwanted exchange of signal between adjacent pixels. The knowledge of this crosstalk and its analyse in relation to the pixel layout and process steps is an important issue for pixel performance improvement. However, crosstalk characterization is quite difficult as it is hard to discriminate signal to crosstalk. A direct way for characterization consists of illuminating only one pixel,[8–10] but as the pixel size decreases, it is harder to optically generate a small enough spot to ensure that only the pixel of interest is excited. So an indirect method needs to be developed for the characterization of pixel crosstalk.

In this paper we present a characterization method based on spectral measurement of pixel response as proposed by Wu[11] but going a step further by coupling this analysis to specific color filter patterns designed to discriminate signal from crosstalk. The goal of these test patterns is to isolate a pixel from its neighborhood (at least in a given spectral range) and also to have un-patterned like pixels. In this objective, we have designed a 5 pixels pattern with one color over the central pixel and all other pixels covered by another color filter.

Depending on the illumination wavelength, the isolated pixel is either the crosstalk-source (the pixel having the largest signal, giving signal by crosstalk to neighbors) or the crosstalk-receiver (the pixel having the smallest signal, receiving signal from neighbors by crosstalk). For instance, a blue pixel surrounded by green pixels is the source of crosstalk in the blue part of the spectrum, and receives the crosstalk in the green wavelengths.

This paper is organized as follow: the section 2 details color filter mask design and the characterization method (setup and data processing). The section 3 details results obtained on 1.4μm pixel. The last section concludes by analyzing the limitations of this method and proposing several ways of improvement.

## 2. CHARACTERISATION METHOD

The goals of this method are to quantify the crosstalk of a pixel and to relate it to the pixel layout and/or process. The simplest solution consists of having a single pixel illuminated and looking at the signal collected by its neighbors. This can be done for instance be using a small spot to illuminate a pixel and if the spot is small enough the intra-pixel sensitivity map can be estimated.[8–10] However, producing a small enough spot becomes arduous for pixel pitch bellow 2μm. One other classical solution consists of masking the surrounding neighbors by metal. This can be done for instance by using a metal level available in the CMOS process but not used in the pixel design. Nevertheless, this solution cannot be used in Back Side Illuminated (BSI) sensors and in Front Side Illuminated (FSI) sensor it can significantly modify the light propagation inside the pixel as diffraction occurs due to the metal window introduced. In addition, due to the additional parasitic capacitance induced, the electrical behavior of the pixel may marginally change.

So we looked for a solution that only introduce a minor change in the pixel structure and that is compatible with very small pixels (below 2μm). The basic idea was to use the color filters in order to mask the neighbors of the pixel of interest. Although this masking is not perfect over the whole spectrum it provides sufficient light isolation over a given spectral band. For instance, if we consider a pixel covered by blue color filter surrounded by pixels covered by green color filter like in figure 1, we can distinguish two behaviors:
- In the blue part of the spectrum, the central pixel presents a large amount of signal compared to its neighbors. Then it is considered as the *crosstalk-source*.
- In the green part of the spectrum, the central pixel present a small amount of signal compared to its neighbors. Then it is considered as the *crosstalk-receiver*.

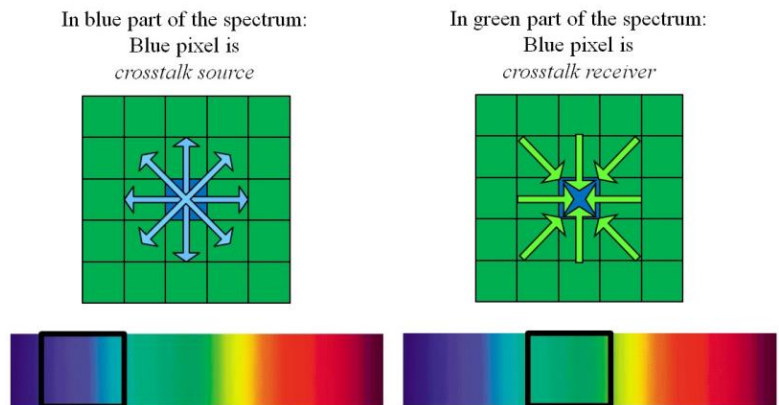

- 
- *Figure 1. Basic principle of pixel isolation using color filter: a pixel covered by blue color filter is surrounded by pixel covered by green color filter.*

The crosstalk can be extracted by a comparison between the spectral response of these pixels and the spectral response of un-patterned pixels (i.e. a group of pixels uniformly covered by a given color filter). On un-patterned pixel, the crosstalk is compensated because all signal send to the neighborhood equals to the signal received from it. So an excess of signal in crosstalk-receiver pixels or lack of signal in crosstalk-source pixel compared to corresponding un-pattern pixel is crosstalk. Because this study is done versus the wavelength, we can spectrally analyze the crosstalk. So we can formulate some hypothesis, related to the pixel layout, about the predominant mechanism as the longer the wavelength, the deeper the electrons are generated.

### 2.1 Mask design

As presented in the previous section we will consider a single pixel, covered by a given color filter surrounding by a uniform field of pixels covered by another color filter (un-patterned). In order to analyze the crosstalk over the full visible spectrum, all possible color combinations must be implemented:

- Blue isolated pixel with green neighbors
- Blue isolated pixel with red neighbors
- Green isolated pixel with blue neighbors
- Green isolated pixel with red neighbors
- Red isolated pixel with blue neighbors
- Red isolated pixel with green neighbors

Then depending on the pixel design, especially for shared transistors architecture, every pattern must be placed above each pixel kind. The figure 2 present the pixel architecture in case of 1T75 structure12 (7 transistors shared between 4 pixels). The pixels are named by the color filter arrangement in Bayer pattern1. So we need to consider 24 different combinations. In order to reduce the noise in measurement, we repeat several time each pattern. This leads to consider the minimal size for a pattern. Another reason for choosing a small pattern is that we must not be significantly different from the standard color deposition process. This will ensure that the filtering is consistent with the standard Bayer one. As the color filters are usually deposited by resist spin coating, the topography of the circuit must be close to the Bayer one. So for the color filter processed in second and third, the pattern of the previous filters and the hole between them must be of the same size (or close). This forbids large area of un-patterned color pixel. We also want to keep the remaining part of the process (mainly the microlens) as close as possible to the standard one.

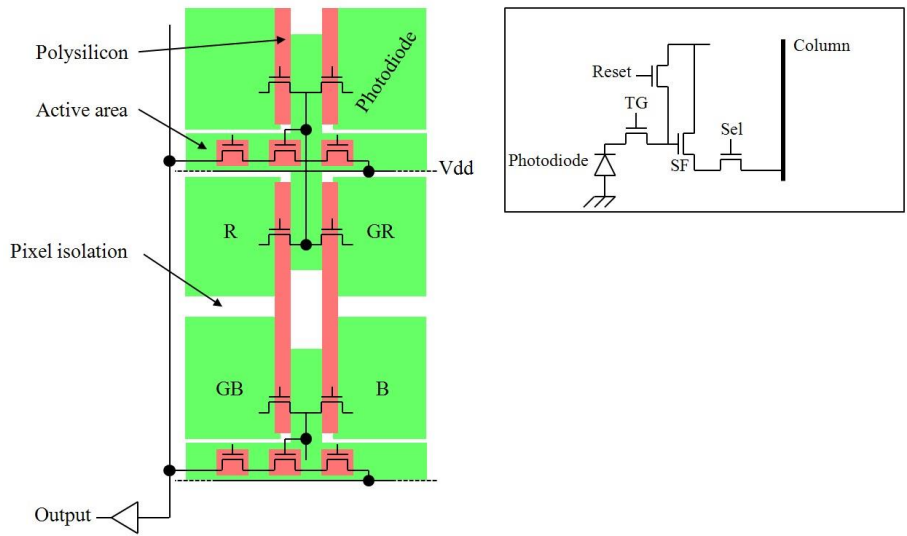

*Figure 2. 1T75 shared pixel architecture: 4 photodiodes share 7 transistors. Inset: 4 transistor active pixel (unshared architecture)*

To define the minimal pattern size we must ensure that for each isolated pixel, direct neighbors (first ring) will only be affected by crosstalk from or toward it. This is why this first ring must be surrounded by pixels (second ring) having the same color. So we need at least 3 rings of pixels around a pixel of interest. This leads to a minimal pattern of 5×5 pixels: two rings around the isolated pixel as shown in figure 3. With this pattern, second ring's pixels can be used to obtain the un-patterned color response. Assuming a test chip of 3Mpix, we can duplicate each pattern (5 × 5 pixels) up to 5000 time (covering the 24 combinations). However on our test chip other experiments have been designed13,14 and implemented, so we only duplicate each pattern about 200 time. The following denomination was chosen in order to identify the pixels inside the pattern (see figure 3):

- *IP*: Isolated Pixel, surrounded by 8 neighbors
- *NN*, *NE*, *EE*, *SE*, *SS*, *SW*, *WW*, *NW*: the *IP*'s neighbors, identified by the direction: North, NorthEast, East, South-East, South, South-West, West and North-West
- *UP*: Un-patterned Pixel: the four corners of the pattern. For these pixels outward crosstalk equals inward crosstalk.

---

[1] R=Red, GR=Green-Red (*i.e.* green filters on the same line than red ones), B=Blue, GB=Green-Blue (*i.e.* green filters on the same line than blue ones)

All these patterns have been implemented and processed on test-chips with under development pixels. Each circuit is made of 3Mpix array and provide raw images. The circuits were characterized on a quantum efficiency (QE) setup described below. Then spectral data have been processed (including pre-processing and averaging) and crosstalk is evaluated from these data.

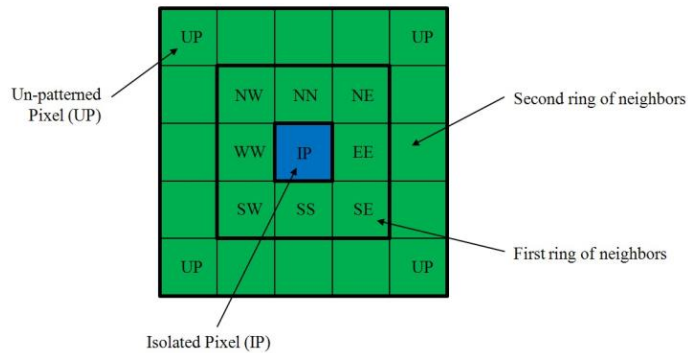

*Figure 3. Patterns used for the crosstalk estimation.*

## 2.2 Characterization bench and data acquisition

The characterization bench is dedicated to spectral response measurement. It is composed by a monochromator producing narrow band illumination (FWHM = 5nm) over the visible range (400nm to 700nm). We scan this range by 10nm step. At the output of this equipment, the light is shaped to produce a uniform illumination, over the chip's sensitive area, using an integrating sphere. The angular distribution of the illumination is controlled by fixing the distance between the sphere output and the sensor (typical value is equivalent to a f-number of 2.8). The irradiance is measured in real time using a second sphere output with similar f-number. The repeatability of the setup has been validated since it was lower than ±2%.

Each characterization run is based on dice coming from the same wafer. Several dice, typically 5, are chosen by rejecting dice with default (faulty pixels, columns or dust, scratch). We ensure that these dice come from the whole wafer surface to be representative of the intra-wafer process dispersion.

For each die, we perform standard quantum efficiency measurement:

- For each wavelength, we acquire several (typically 10) illuminated images. The integration time is chosen so that the signal level of the most sensitive pixel at this wavelength is close to half of the saturation level.
- Several (typically 10) images in darkness are acquired, with the shortest integration time used over the spectrum. As the dark current in room temperature is very low, the integration time do not have a significant impact on results.

## 2.3 Data processing

### 2.3.1 Data preprocessing

Using the previous images, we apply the standard pre-processing steps:

- For each wavelength, average the images in order to reduce temporal noise. Do as well for the dark images
- Subtract, pixel by pixel, the average dark image to the average image under light, for each wavelength, this suppress the dark signal
- Extract the signal of each kind of pixel (IP, UP, NN, NE, etc)for all patterns and average the pixels of the same kind to reduce the spatial noise (non-uniformity)
- In addition the dispersion of results is checked before using the mean value.

### 2.3.2 Crosstalk evaluation

Crosstalk estimation is based on comparison between isolated pixel, first ring of surrounding pixels (NN, NE, EE, etc) and un-patterned pixels. Figure 4 illustrates the example of a blue pixel. The difference between UPblue the response of the eight pixel surrounding a green isolated pixel gives the signal lost in the blue part of the spectrum (going toward the green central pixel) and, in the green part of the spectrum, the excess of signal (coming from the green central pixel). This can be analyzed either by direct subtration between QE curves or by normalizing this difference by a reference signal. Once may choose the UP signal as reference as it is the crosstalk free signal, so it is the best signal we can expect from the pixel. Nevertheless, this normalization is not obvious as it can only be done for a spectral range around the peak response of the color filter: otherwise, the reference signal is too low to provide accurate normalization.

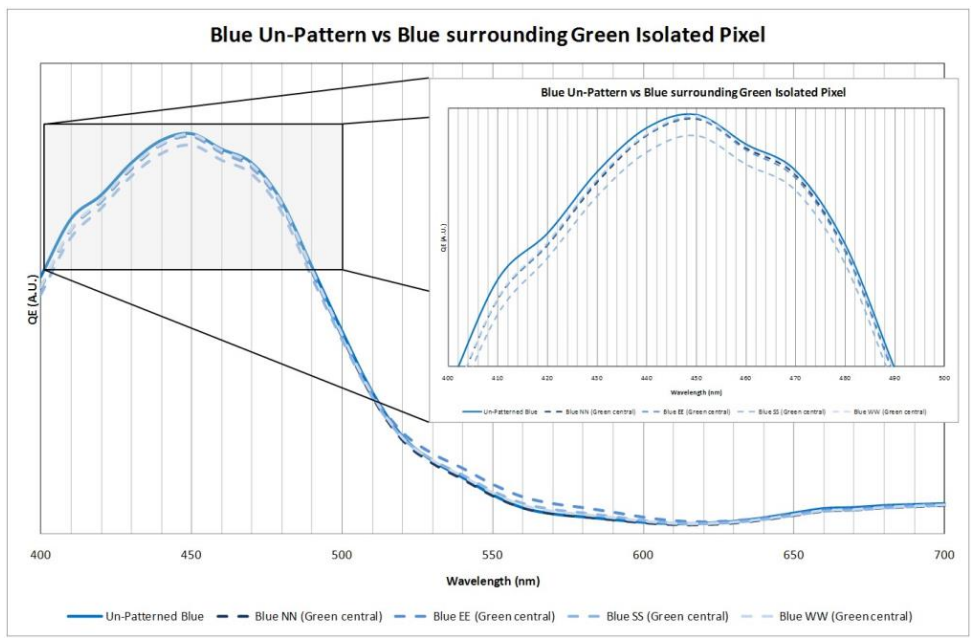

*Figure 4. QE curves of blue Un-patterned Pixel (UP$_{Blue}$) and blue pixels surrounding a green Isolated Pixel (IP).*

So we analyze the result directly as QE points by subtracting the surrounding pixel signal to the UP signal. From 1.4µm pixel, the diagonal crosstalks are negligible. So, for each pixel of interest we define the 4 main directions of crosstalk based on the layout as shown in figure 5. The typical output is then given by the figure 6 for a Green IP surrounded by

Blue pixels (extracted from curves shown in figure 4). We can clearly see the additional signal coming from the blue pixels in the blue part of the spectrum and the loss toward these pixels in the green part.

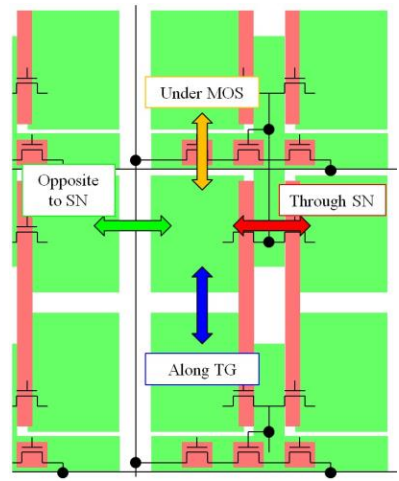

Figure 5. Main crosstalk directions, defined with respect to the layout: horizontal crosstalk (Through Sense Node (SN) and Opposite to SN) and vertical crosstalk (Along Transfer Gate (TG) and under MOS).

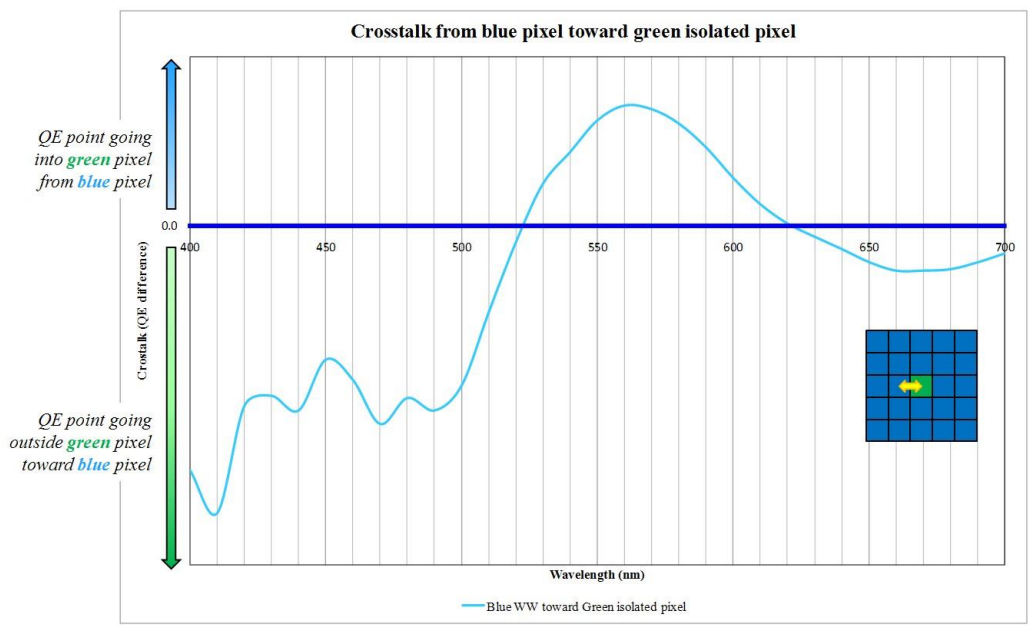

Figure 6. Crosstalk extracted from QE curves shown in figure 4.

## 3. RESULTS

### 3.1 Crosstalk

The method presented here gives access to crosstalk and its analyze ywith respect to the pixel layout. For instance the figure 7 shows the crosstalk extracted from Blue IP surrounded by green pixels. We can see a main direction in the blue part of the spectrum: along the TG (Transfer Gate). Other directions exhibit low crosstalk and no significant differences: the surface isolation is efficient in these directions. The case of along TG direction has been analyzed using electromagnetic simulations15,16 as an optical coupling through the polysilicon of the transfer gate which is shared by two pixels. The other directions have very low level of crosstalk for short wavelengths. Indeed, the photons are absorbed very closely to the silicon surface, so the topography of the pixel at silicon surface play an important role. For longer wavelengths, two behaviors are clearly present. For the directions opposite to the sense node and under the MOS transistors, the crosstalk is small because the pixel isolation ensures a good separation between pixel. As the pixel isolation is not complete along the sense node side, some electrons can drift under the sense node toward the direct neighbor. That is why the crosstalk through the sense node is high. Also, the longer the wavelength, the higher the diffraction by the poly line of the transfer gate. Consequently, the crosstalk along the transfer gate is still high.

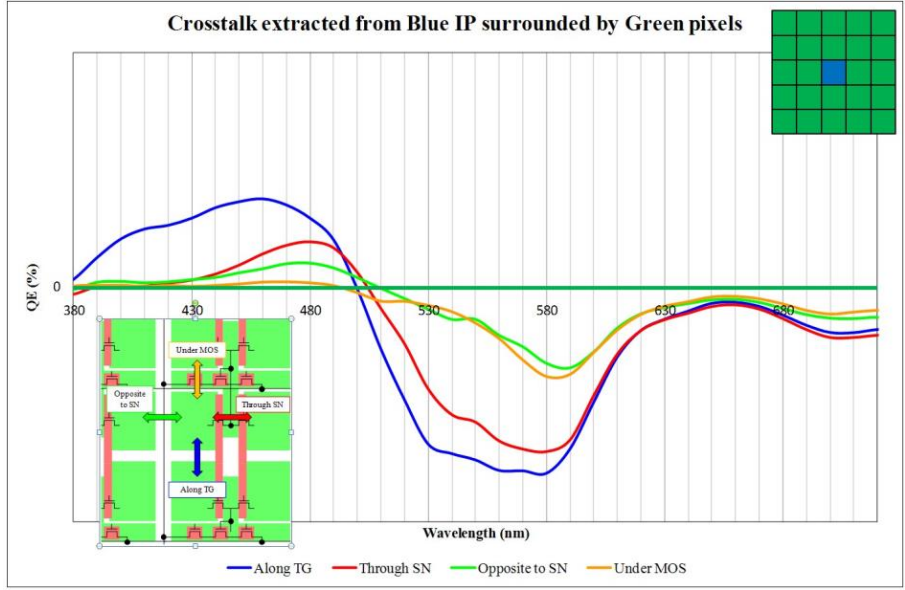

*Figure 7. Crosstalk extracted for a Blue isolated pixel IPBlue surrounded by un-patterned Green pixels. The Blue un-patterned UPBlue response is take as reference.*

By changing the pattern, we have access to another part of the spectrum. For instance, considering a Red isolated pixel IPRed surrounded by green pixel as shown in figure 8 allows the estimation of crosstalk at long wavelengths. With this pattern we see that crosstalk along transfer gate and through sense node remains the highest as the root cause is still the same. On the red part of the spectrum, we can see that crosstlak in the direction opposite to the sense node is twice higher than under the MOS transistor. This is due to the double isolation in the latter direction. It increases the length that electrons have to cross over by diffusion, compared to the opposite to SN direction.

### 3.2 Comparison with Bayer patterned pixel

An interesting comparison is between standard Bayer QE curves and the three un-patterned QE curves. In figure 9, the Bayer QE curves are in solid lines and the un-patterned ones in dotted lines. This shows clearly the impact of crosstalk and Bayer patterning on this 1.4µm pixel: reduction of peak signals and worse rejections.

We can go deeper in the analysis by reconstructing the Bayer QE curves from the crosstalk we have measured. In fact, the crosstalk is known over the whole spectrum and for any position. So we can apply it on UP response. Then we will compare these synthetic responses to the real response of Bayer patterned pixels. The calculation of the reconstructed Bayer curves $\widetilde{R}, \widetilde{GR}, \widetilde{GB}, \widetilde{B}$ from the UP curves (UPR, UPGR, UPGB, UPB) and the horizontal/vertical crosstalk

(XR→GR, XR→GB, XGR→R, XGR→B, XGB→R, XGB→B, XB→GR and XB→GB) is given by the following equations. We use simply subtract outward crosstalk and add inward crosstalk to each UP curve:

$$\widetilde{R} = (UP_R - 2X_{R \to GR} - 2X_{R \to GB}) + (2X_{GR \to R} + 2X_{GB \to R})$$
$$\widetilde{GR} = (UP_{GR} - 2X_{GR \to R} - 2X_{GR \to B}) + (2X_{R \to GR} + 2X_{B \to GR})$$
$$\widetilde{GB} = (UP_{GB} - 2X_{GB \to R} - 2X_{GB \to B}) + (2X_{R \to GB} + 2X_{B \to GB})$$
$$\widetilde{B} = (UP_B - 2X_{B \to GR} - 2X_{B \to GB}) + (2X_{GR \to B} + 2X_{GB \to B})$$

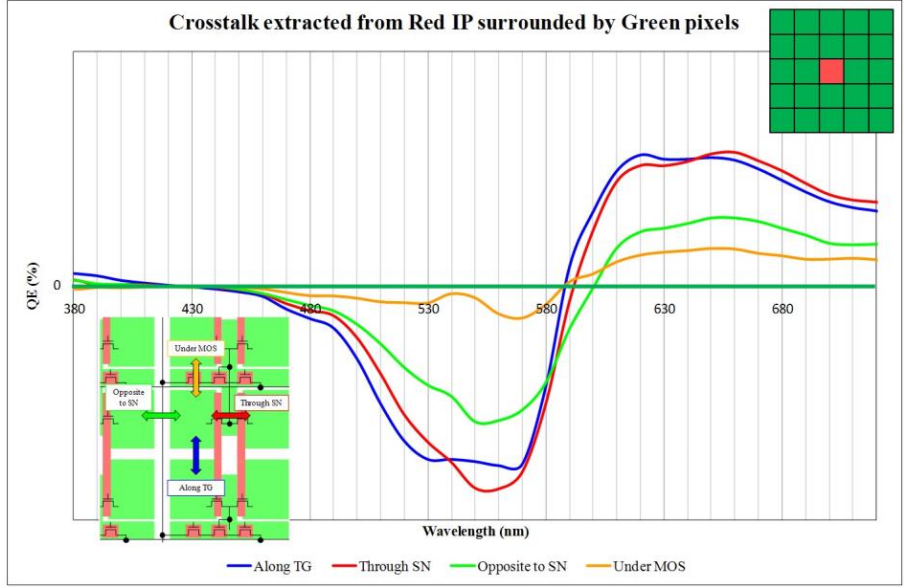

*Figure 8. Crosstalk extracted for a Red isolated pixel IP$_{Red}$ surrounded by Green pixels. The Red un-patterned UP$_{Red}$ response is take as a reference.*

We obtain the curves shown in the figure 9 (to simplify, we only present the Green-Red pixel response as the Green-Blue one is very close).

This acts as a self validation of the measurement. As dashed lines should be close to solid ones. We can clearly see that our method accurately estimate the crosstalk for wavelengths shorter than 540nm, but for longer wavelengths it is not properly estimated. The exchange with the Red pixel is underestimated as the peak response of the reconstructed Red is overestimated and the Blue and Green-Red signal are underestimated. The root cause of this discrepancy has to be investigated. However we can tighten a little the possible causes. The underlying pixel (below color filter) structure may not be involved as it is unchanged. So we can suspect an optical effect. As the response in the short wavelengths is correct for all color, the filter thickness in UP and IP pixels is close to the thickness in Bayer pattern and can be cleared. So it could be a process effect: the photolithography step of the color filter varies with the pattern density, so isolated pixel will not behave as the Bayer ones. This leads to some diffraction effect that depend on the pattern (and so change the light distribution inside the silicon). This point appears as the main limitation of our method. We may also mention the diagonal crosstalk between Red ang Blue pixels that are not well considered here. Nevertheless data don't show significant level of diagonal crosstalk even at long wavelengths.

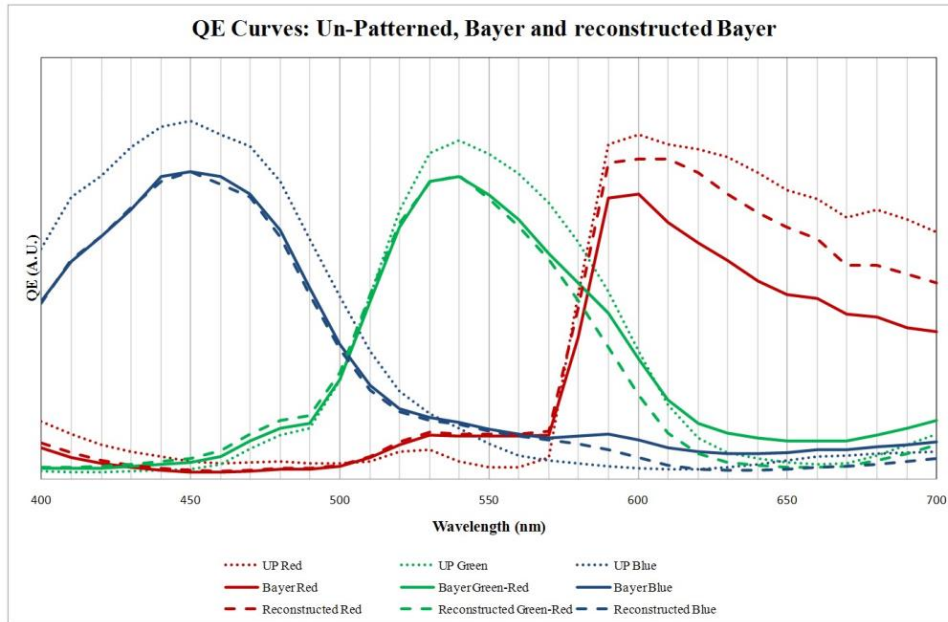

*Figure 9. Reconstructed Bayer curves (in dash lines) from UP (in dotted lines) and crosstalk curves versus the true Bayer QE curves (in solid lines).*

## 4. PERSPECTIVES AND CONCLUSION

The methodology presented in this paper allows to estimate the crosstalk on very small pixel where direct methods (like small spot illumination) are not available. By designing a specific color filters mask-set and by measuring the quantum efficiency, we are able to extract several information: the crosstalk free QE, the amount crosstalk versus the wavelength and the pixel layout. We also see some impacts of Bayer patterning that cannot be predicted from the previous analysis.

However, we can think about several improvements of this method. At first the use of a black color filter will directly give access to the crosstalk over the whole spectrum. The analysis of Bayer pattern has also to be deepen to understand the cause of departure between real responses and the ones reconstructed from crosstalk analysis. This will be a great step toward the modelization of the crosstalk in small pixel allowing to simulate the pixel response for any un-patterned color filter transmission curve (such as one obtained on glass wafers).